\documentclass[aps,prl,twocolumn,superscriptaddress,reprint,showpacs]{revtex4-2}
\usepackage{graphicx}
\usepackage{setspace}
\usepackage{amsmath}
\usepackage{amssymb}
\usepackage{color}
\usepackage{array}
\usepackage{subfigure}
\usepackage{hyperref}
\usepackage{float}
\usepackage{lipsum}

\usepackage[all]{xy}
\newcommand{\RN}[1]{%
  \textup{\uppercase\expandafter{\romannumeral#1}}
}

\begin{document}
\title{Refine coherent control of atomic qubits via wave-function approach conditioned on no-decay}

\author{Yuan Sun}
\email[email: ]{yuansun@siom.ac.cn}
\affiliation{Shanghai Institute of Optics and Fine Mechanics, Chinese Academy of Sciences, Shanghai 201800, China}
\affiliation{University of Chinese Academy of Sciences, Beijing 100049, China}

\begin{abstract}
As a fundamental phenomenon in quantum systems, spontaneous emission constitutes an inevitable source of error, which ultimately degrades the fidelity of quantum logic gates. A successful quantum logic gate needs to operate on the condition that no decay event, such as spontaneous emission, occurs. Such successes can be ensured by post-selection based on syndrome extraction  according to the theory of quantum error correction or quantum error mitigation. In this case, the wave function of qubits remains a pure state but is subject to additional influences from spontaneous emission, even without actual decay events. Therefore, such a process must be appropriately described by a modified version of Schr\"odinger equation for the dynamics conditioned on no-decay. Calculations reveal that this effect must be seriously taken into consideration for the design of high-fidelity quantum logic gates. With respect to realistic experimental conditions, even if the coherence is well preserved, improving the fidelity of manipulating physical qubits requires careful consideration of the subtle influences of decay processes such as spontaneous emission. Specifically, the gate and readout processes in the atomic qubit platform are discussed.
\end{abstract}
\maketitle

The research of quantum computing exemplifies how principles of quantum mechanics can unlock unprecedented information-processing capabilities. The quantum advantage and quantum utility are transitioning from theoretical concepts to practical performance improvements \cite{Preskill2012paper, Wujunjie2018NSR, Google2019Sycamore, Nature618.500, GoogleRQCnature2024, Google2024Willow, Saffman2025LogiQ_arXiv}. Fundamentally, the road map of quantum computing originates in the Hilbert space spanned by two-qubit register states $\{|0\rangle, |1\rangle\}$. It inherently imposes the requirement that successful quantum logic gates must maintain population within this space. Realistically, the physical qubit's $\{|0\rangle, |1\rangle\}$ always have finite limited coherence time, and very often the quantum logic gates require coherent transitions or couplings of populations to states outside $\{|0\rangle, |1\rangle\}$. Therefore, the population suffers loss in various ways that eventually lead to dissipation through spontaneous emission, resulting in irreversible population leakage. For instance, in the atomic qubit platform \cite{PhysRevA.62.052302, RevModPhys.82.2313, Saffman_2018NSR}, the Rydberg blockade effect \cite{nphys1178, PhysRevLett.102.013004, PhysRevLett.104.010503, Yuan2024SCPMA} plays an essential role in enabling the high-fidelity two-qubit gates \cite{PhysRevApplied.13.024059, PhysRevApplied.15.054020, PhysRevA.105.042430, Lukin2023nature, Yuan2025SWAP, Infleqtion2025PRXquantum, AtomComputing2025arXiv}, and thus the coherent ground-Rydberg transition becomes indispensable \cite{PhysRevA.97.053803, PhysRevApplied.15.054020, YouLi2024Optica, Yuan2025CPL, Infleqtion2025PRXquantum}. Therefore, there is a non-negligible probability that some population will not return to the qubit register state space due to the inherent physical complexity and experimental imperfections of the ground-Rydberg transition. 

While the spontaneous emission and other possible decoherence sources continuously affect the physical qubit, both noisy intermediate-scale quantum (NISQ) and fault-tolerant quantum computing (FTQC) demand that a successful quantum logic gate operates under the condition that no decay event occurs. If the population leakage surpasses a certain level, it will be recognized as an error by the quantum error correction (QEC). During the successful coherent control, although the wave function of the physical qubits stays as a pure state, its time evolution deviates from the usual Schr\"odinger equation because the entire process is conditioned on no-decay. In the atomic qubit platform, beyond typical single-qubit and Controlled-PHASE gates, a proposed solution for high-speed, high-connectivity large-scale arrays is the buffer-atom framework, which includes the buffer-atom-mediated (BAM) gate \cite{Yuan2024SCPMA, Yuan2024FR2} and Rydberg blockade SWAP gate \cite{Yuan2025SWAP} as key components. The designs of these gates need to carefully consider the nuances brought by the time evolution conditioned on no-decay. 

When conditioned on no-decay, although the wave function remains a pure state, it is subject to additional influences from decay channels such as spontaneous emission, as it is constantly probed by the environment. This also connects with post-selection of quantum trajectories in the Liouvillian formalism \cite{Murch2019NP, PhysRevA.100.062131, PhysRevA.101.062112} for open quantum systems with constant driving, steady state, well-defined eigenvalues, adiabatic evolution, or statistical properties \cite{PhysRevB.102.201103, PhysRevLett.127.140504, PhysRevA.107.022216, PhysRevB.110.094315, PhysRevB.111.024303}, which has been extensively used in studying exceptional points (EP). 

\begin{figure}[h]
\centering
\includegraphics[width=0.46\textwidth]{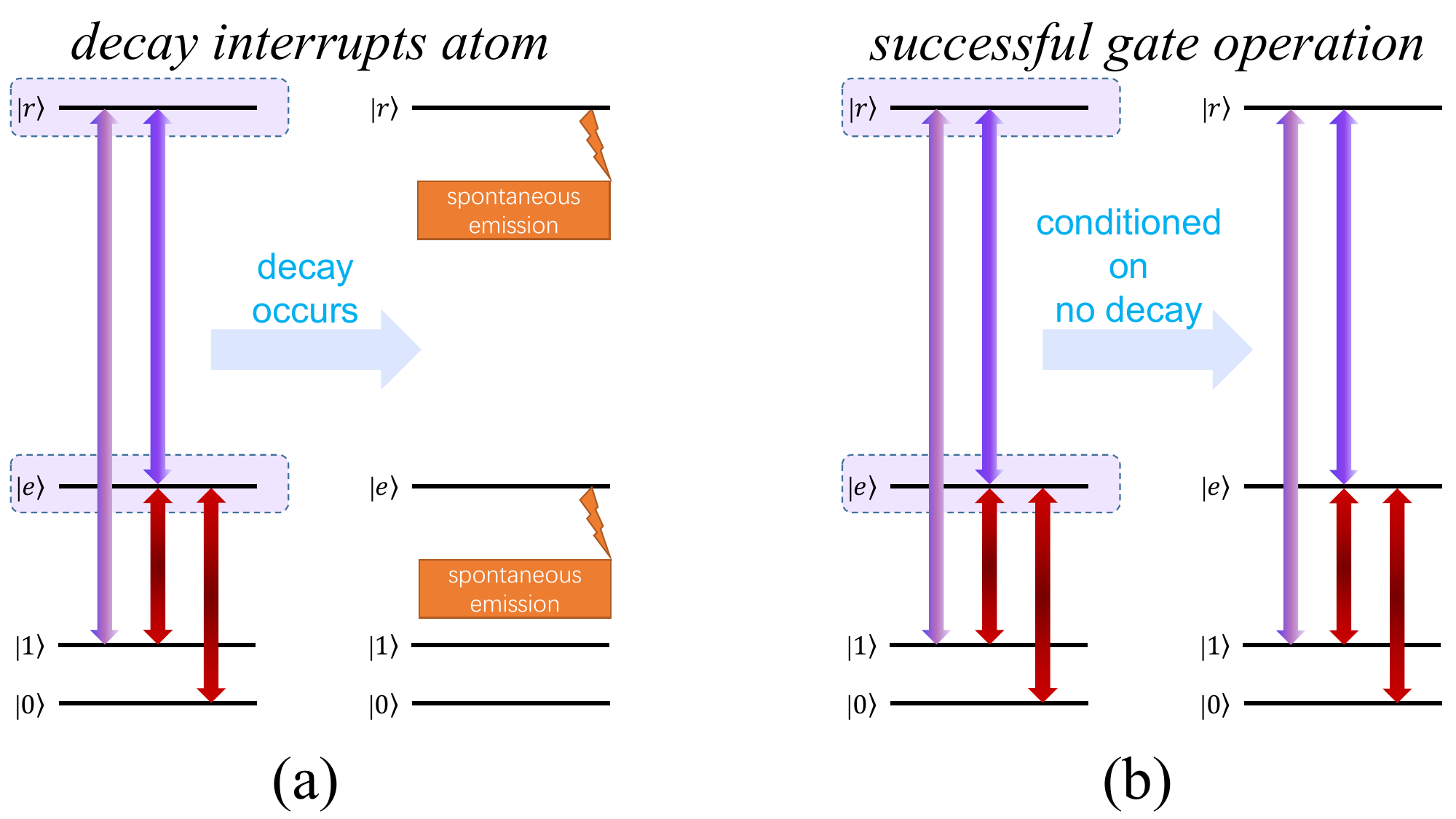}
\caption{(Color online) Controlling atomic qubits requires atom-laser interaction involving atomic levels subject to decay. For example, the intermediate state $|e\rangle$ and Rydberg state $|r\rangle$ are subject to spontaneous emissions. A successful gate operation demands that decay does not take place.}
\label{fig:no_decay}
\end{figure}

Therefore, it is worthwhile to study the time evolution of physical qubits conditioned on no-decay when the wave function receives time-dependent drivings. According to the Monte-Carlo wave function (MCWF) approach \cite{PhysRevLett.68.580, Molmer:93JOSAB, PhysRevLett.70.2273, RevModPhys.70.101, Wiseman_book}, in a quantum trajectory where no quantum jump takes place and the wave function remains a pure state, a special form of Schr\"odinger equation conditioned on no-decay (SECOND) governs the dynamics. The rest of the content is organized as follows. First, we construct the equations of motion via SECOND, and subsequently analyze the influences on atomic qubits, with emphasis on the quantum logic gates as shown in Fig. \ref{fig:no_decay}. Then we discuss MCWF conditioned on no-decay for practical detection processes and the potential implications for ongoing experimental efforts.

We first develop the abstract form of SECOND for the wave function $\mathbf{C} = [C_1, C_2, \ldots, C_N]^T$ of an $N$-state system. Without considering the decay effects, suppose the unitary time evolution is governed by the Hamiltonian $\mathcal{H}_\text{U}$ such that $i\hbar\frac{d}{dt}\mathbf{C}(t) = \mathcal{H}_\text{U}(t)\mathbf{C}(t)$ with $\mathcal{H}_\text{U}^\dagger=\mathcal{H}_\text{U}$. Now consider the decay effects as $M$ channels involving $C_{e_1}, C_{e_2}, \ldots, C_{e_M}$ with decay rates $\gamma_{e_1}, \gamma_{e_2}, \ldots, \gamma_{e_M}$, respectively. Define $\mathcal{H}_\text{D}$ and $\mathcal{H}_\text{R}$ as:
\begin{eqnarray}
\label{eq:H_D}
{\mathcal{H}_\text{D}}_{jk}=-i\hbar\sum_{m=1}^{M}\frac{\gamma_{e_m}}{2}\delta_{je_m}\delta_{e_mk},\\
\label{eq:H_R} 
{\mathcal{H}_\text{R}}_{jk}=i\hbar\sum_{m=1}^{M}\frac{\gamma_{e_m}}{2}|C_{e_m}|^2\delta_{jk}. 
\end{eqnarray}
Within the MCWF framework, post-selection or conditioning on no-decay ultimately indicates that decay does not happen for each infinitesimal time step $dt$ of the evolution. Therefore, for non-pathological cases where the relevant channels are well described by the exponential decay, the following relation now establishes the SECOND $i\hbar\frac{d}{dt}\mathbf{C}(t) = \mathcal{H}_S(t)\mathbf{C}(t)$: 
\begin{equation}
\label{eq:SECOND_general}
\mathcal{H}_S = \mathcal{H}_\text{U} + \mathcal{H}_\text{D} + \mathcal{H}_\text{R},
\end{equation}
where the subscripts U, D, and R represent unitary time evolution, decay, and renormalization of the wave function, respectively. This framework applies to a standalone atom as well as interacting atoms, provided that $\mathcal{H}_\text{U}$ includes the interaction terms for a multi-atom system, and $\mathcal{H}_\text{D}$, $\mathcal{H}_\text{R}$ are appropriately adjusted for multi-atom states. Obviously, it is non-linear and non-Hermitian, and practical applications may require numerically solving the time-dependent differential equation system. 

Consider the basic two-level atom model as an example. Suppose the wave function is $[X(t), Y(t)]$ with decay rates $\gamma_x, \gamma_y$. The atom-laser interaction is time-dependent with Rabi frequency $\Omega(t)$ and detuning $\Delta(t)$. Then its SECOND is the following:
\begin{subequations}
\label{eq:SE_no-decay_2level}
\begin{eqnarray}
i\dot{X} = \big(-\frac{i}{2}\gamma_x+ \frac{i}{2}(\gamma_x|X|^2+\gamma_y|Y|^2)\big)X + \frac{\Omega}{2} Y, \\
i\dot{Y} = \frac{\Omega}{2} X + \big(\Delta  - \frac{i}{2}\gamma_y + \frac{i}{2}(\gamma_x|X|^2+\gamma_y|Y|^2)\big)Y,
\end{eqnarray}
\end{subequations}
where $\gamma_x|X|^2+\gamma_y|Y|^2$ stems from the renormalization of wave function after decoherence. 

The conservation of probability in Eq. \eqref{eq:SECOND_general} needs verification and it suffices to check the two-level case using Eq. \eqref{eq:SE_no-decay_2level}. Let $\xi(t)=|X(t)|^2+|Y(t)|^2-1$, and then it reduces to proving $\dot{\xi}(t) \equiv 0$. Arithmetic calculations from Eq. \eqref{eq:SE_no-decay_2level} yield that:
\begin{equation}
\label{eq:SECOND_unitary_condition}
\dot{\xi} = (\gamma_x|X|^2+\gamma_y|Y|^2)\xi.
\end{equation}
Provided $\xi(0)=0$, Eq. \eqref{eq:SECOND_unitary_condition} shows that all finite-order derivatives of $\xi$ are zero at $t=0$, which suggests that $\xi=0, \forall t$.

With these preparations, we will analyze the influence of Eq. \eqref{eq:SECOND_general} on quantum computing, with emphasis on the atomic qubit platform. The discussions will focus on the implications for designing high-fidelity single-qubit and two-qubit gates. 

For atomic qubits, the local single-qubit gate \cite{PhysRevLett.114.100503, PhysRevLett.121.240501, Infleqtion2025PRXquantum} typically operates via the three-level Raman transition. Previous designs have generally regarded decay as an inherent ingredient but have not yet evaluated the nuances of conditional no-decay. According to Eq. \eqref{eq:SECOND_general}, the time evolution conditioned on no-decay of $\mathbf{C} = [C_0, C_1, C_e]^T$ obeys $i\hbar\frac{d}{dt}\mathbf{C} = \mathcal{H}_3\mathbf{C}$ with $\mathcal{H}_3/\hbar$ as:
\begin{equation}
\label{eq:Raman_H3}
\begin{bmatrix} i\frac{\gamma}{2}|C_e|^2 & 0 & \frac{\Omega_0(t)}{2}\\ 
0 & \delta(t) + i\frac{\gamma}{2}|C_e|^2 & \frac{\Omega_1(t)}{2}\\ 
\frac{\Omega_0(t)}{2} & \frac{\Omega_1(t)}{2} & \Delta(t) -i\frac{\gamma}{2} +i\frac{\gamma}{2}|C_e|^2 \end{bmatrix},
\end{equation}
with $\Delta, \delta$ being single-photon and two-photon detunings respectively, and $\gamma$ being state $|e\rangle$'s decay rate.

Consider a $\pi/2$-pulse corresponding to the Hadamard gate. Express the waveform function $f$ as truncated Fourier series $[a_0, a_1, \ldots, a_N]$, which represents $f(t)=2\pi \times \big(a_0 + \sum_{n=1}^{N} a_n\exp(2\pi i nt/\tau) + a^*_n\exp(-2\pi i nt/\tau) \big)/(2N+1) \text{ MHz}$ with reference time $\tau = 1\, \mu\text{s}$ here. The pulse waveform is a variant of the Gaussian function, suppressing high-frequency components and zeroing the end points \cite{PhysRevApplied.20.L061002}. The parameters include $\Omega_0=\Omega_1$ given by $[208.05, -92.59, -3.66, -3.35, -2.38, -2.03]$, $\Delta=2\pi\times 1\text{ GHz}$, $\delta=0$, designed for the unitary time evolution.

\begin{figure}[t!]
\centering
\includegraphics[width=0.4\textwidth]{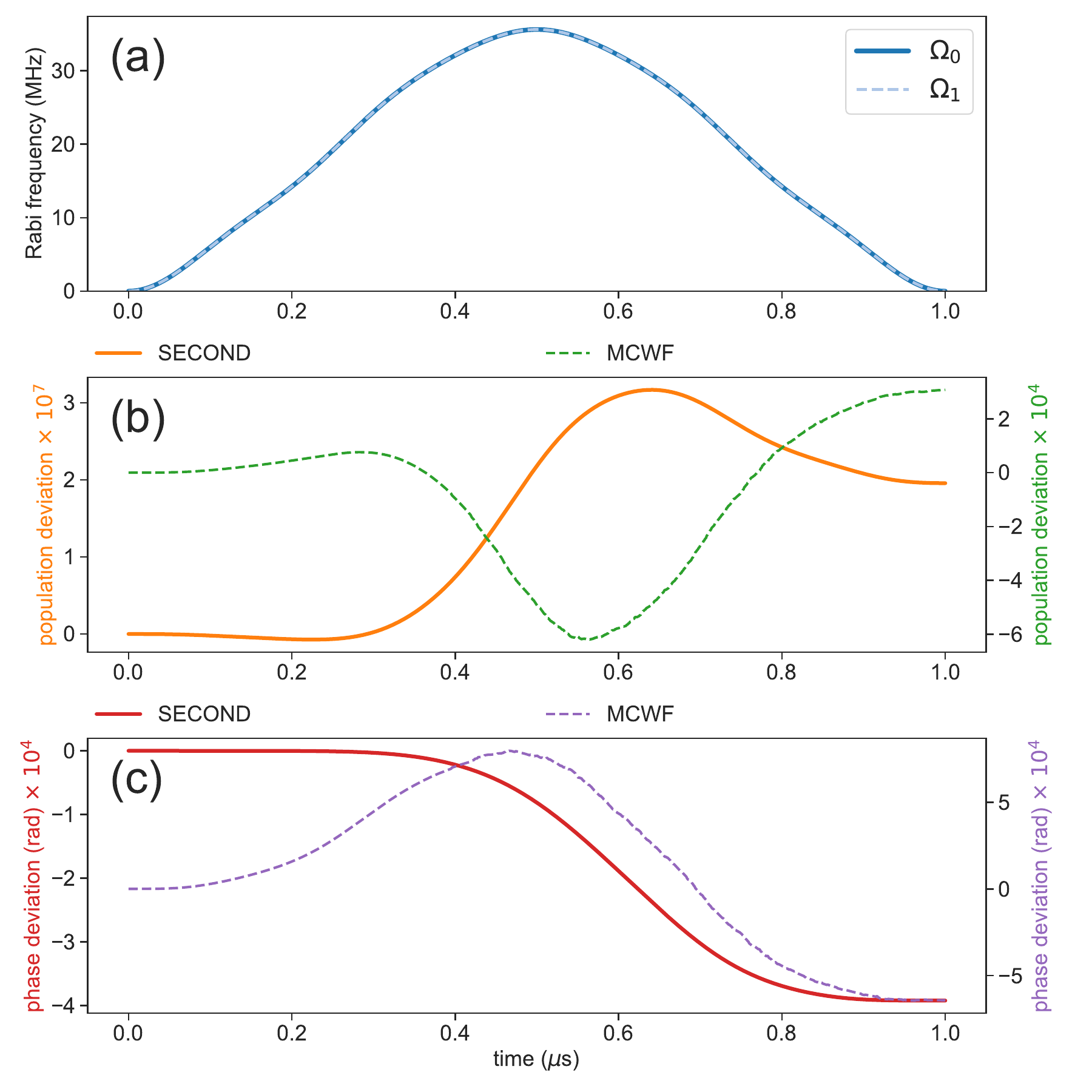}
\caption{(Color online) Comparisons between unitary time evolution ($U$), SECOND ($S$) and MCWF ($M$, averaged over 250000 trajectories). (a) Waveforms of $\Omega_0(t), \Omega_1(t)$. (b) Population differences $P_S - P_U$ and $P_M - P_U$ in state $|0\rangle$. (c) Phase differences $\phi_S - \phi_U$ and $\phi_M - \phi_U$ in state $|0\rangle$.}
\label{fig:single_qubit_gate}
\end{figure}

The wave function is then numerically evaluated with the result shown in Fig. \ref{fig:single_qubit_gate}, where $\gamma=2\pi\times 1\text{ MHz}$ and the MCWF calculation assumes that $|e\rangle$ decays to $|0\rangle, |1\rangle$ with equal probabilities. Not surprisingly, the time evolution conditioned on no-decay, as governed by Eq. \eqref{eq:Raman_H3}, exhibits clear distinctions compared with the unconditional case. The difference in the population is on the order of the average spontaneous emission probability $p_\gamma$ over the interaction. In particular, both the amplitude and phase of the wave function are altered due to conditional no-decay.

\begin{figure}[h]
\centering
\includegraphics[width=0.4\textwidth]{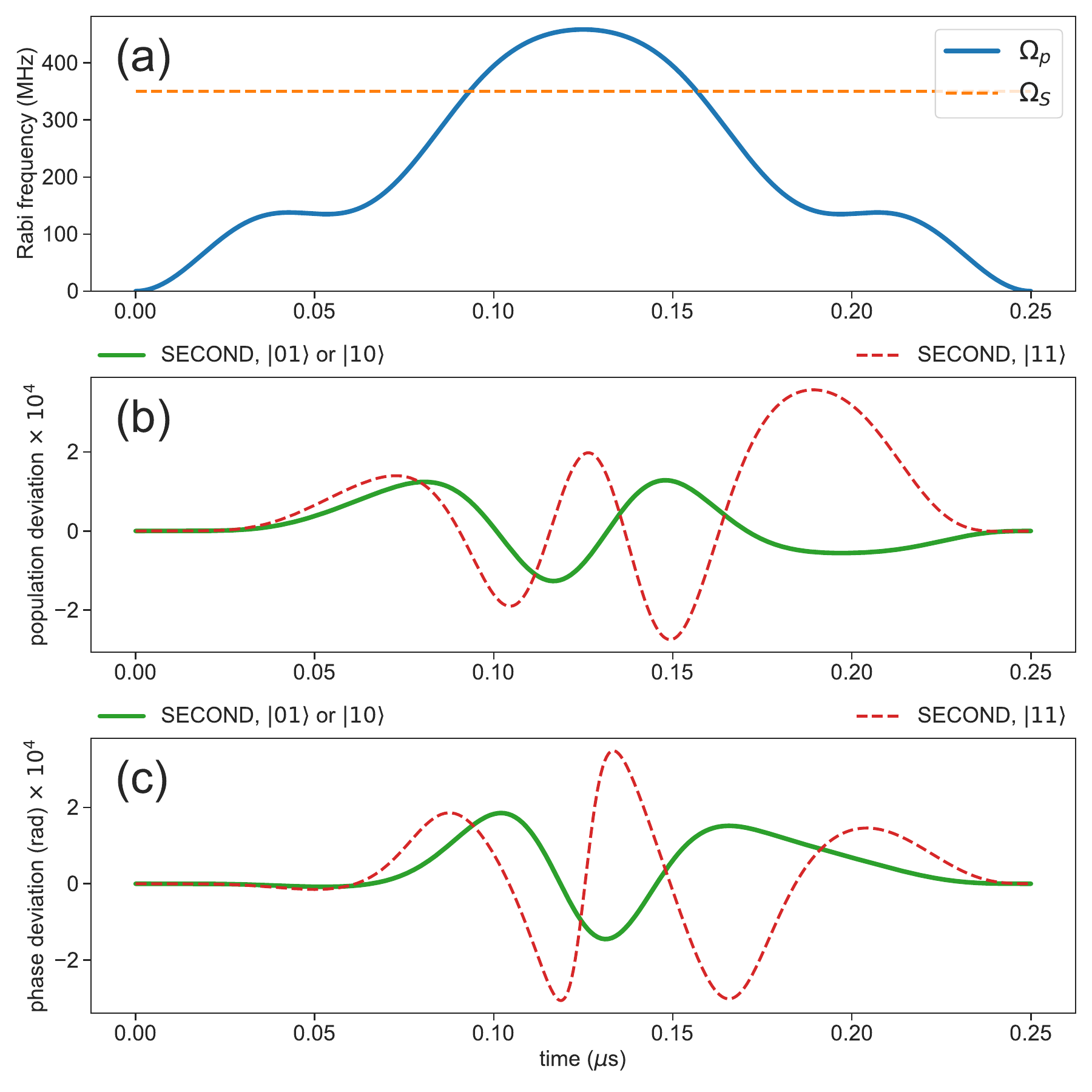}
\caption{(Color online) Comparisons between unitary time evolution ($U$) and SECOND ($S$) for the CZ gate process. (a) Waveforms of $\Omega_p(t), \Omega_S(t)$. (b) Population differences $P_S - P_U$. (c) Phase differences $\phi_S - \phi_U$.}
\label{fig:two_qubit_gate}
\end{figure}

We now study the two-qubit gates and in particular the CZ gate with the well-established method of synthetic modulated driving. Fig. \ref{fig:two_qubit_gate} presents a typical result with the reference time $\tau = 0.25\, \mu\text{s}$ and $\Omega_p$ given by $[2338.8, -1082.4, 230.9, -176.6, -138.9, -2.3]$. Other parameters include: $\Omega_S = 2\pi\times 350 \text{ MHz}$, $\Delta = 2\pi\times 5 \text{ GHz}$, $\delta = 2\pi\times -0.53 \text{ MHz}$, and the Rydberg blockade strength $B = 2\pi\times 100 \text{ MHz}$. Here the decay rates $\gamma_e, \gamma_r$ of intermediate and Rydberg states are set as $2\pi\times 5\text{ MHz}$ and $2\pi\times 10\text{ kHz}$ respectively.

The unitary evolution and SECOND calculations agree well with each other at the end times of the gate in Fig. \ref{fig:two_qubit_gate}, and this agreement is not coincidental. A perturbation analysis can help to interpret this observation. Given $\mathbf{C}_0$ as the unperturbed wave function solution satisfying the unitary time evolution $i\hbar\frac{d}{dt}\mathbf{C}_0(t) = \mathcal{H}_\text{U}(t)\mathbf{C}_0(t)$, suppose $\mathbf{C}_\epsilon(t)$ is the correction such that $i\hbar\frac{d}{dt}\big(\mathbf{C}_0(t)+\mathbf{C}_\epsilon(t)\big) = \big(\mathcal{H}_\text{U}(t)+\mathcal{H}_\text{D}(t) + \mathcal{H}_\text{R}(t)\big)\big(\mathbf{C}_0(t)+\mathbf{C}_\epsilon(t)\big)$ with $\mathcal{H}_\text{D}(t) + \mathcal{H}_\text{R}(t)$ as the perturbation, and neglect higher order terms in seeking values of $\mathbf{C}_\epsilon(0), \mathbf{C}_\epsilon(T)$. Then $\mathbf{C}_\epsilon(t)$ satisfies:
\begin{equation}
(i\hbar\frac{d}{dt} - \mathcal{H}_\text{U})\mathbf{C}_\epsilon = (\mathcal{H}_\text{D} + \mathcal{H}_\text{R})\mathbf{C}_0.
\end{equation}
For a phase gate on $[0, T]$, the amplitude of the modulated driving generally smoothly reduces to zero at the end points $t=0, T$ \cite{PhysRevApplied.20.L061002}, so that effectively $\frac{d}{dt}\mathbf{C}_\epsilon(0) = \frac{d}{dt}\mathbf{C}_\epsilon(T) = 0$. A good gate design tends to avoid leaving populations on lossy states, which indicates $(\mathcal{H}_\text{D} + \mathcal{H}_\text{R})\mathbf{C}_0|_0 = (\mathcal{H}_\text{D} + \mathcal{H}_\text{R})\mathbf{C}_0|_T = 0$. Therefore, it implies that $\mathbf{C}_\epsilon(0)=\mathbf{C}_\epsilon(T)=0$, which corroborates Fig. \ref{fig:two_qubit_gate}.
This observation applies to single-, two- and multi-qubit phase gates with appropriate modulated drivings. More advanced and complicated Rydberg blockade two- and multi- qubit phase gates also possess similar features when conditioned on no-decay, such as the BAM and heralded self-correction gates \cite{Yuan2024SCPMA, Yuan2024FR2, Yuan2025CPL}. 

\begin{figure}[h]
\centering
\includegraphics[width=0.4\textwidth]{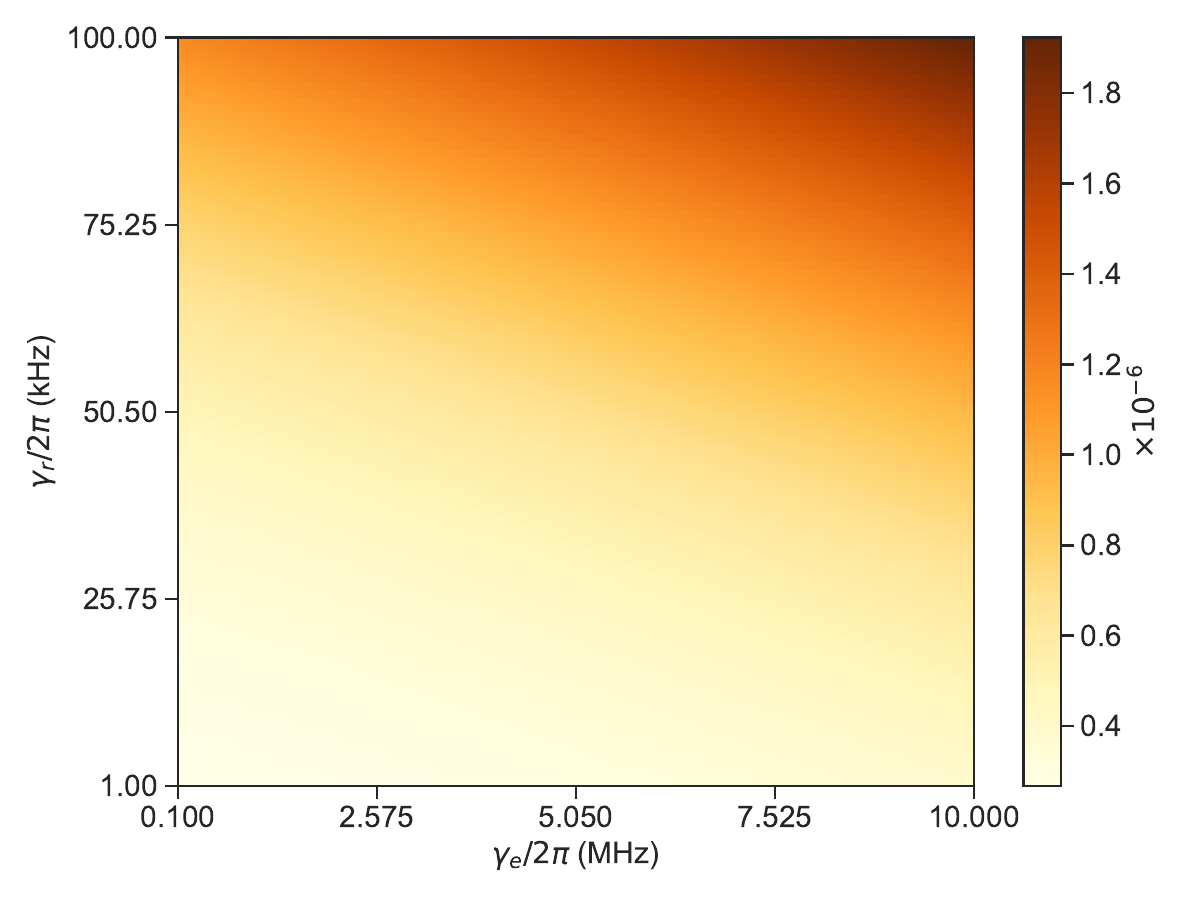}
\caption{(Color online) Gate errors $\varepsilon=1-f$ for CZ gate operation conditioned on no-decay, calculated via SECOND. The values of $\gamma_e, \gamma_r$ are scanned. }
\label{fig:gate_error_scan}
\end{figure}

The influence of decay on gate operation conditioned on no-decay merits further investigation. In particular, how the gate fidelity $f$ varies with changes in $\gamma_e, \gamma_r$ can be numerically examined. A typical result is shown in Fig. \ref{fig:gate_error_scan}, using the same modulated driving as in Fig. \ref{fig:two_qubit_gate}. Over a relatively wide range of decay rates, the change in gate fidelity remains at a level of $\sim 10^{-6}$ when the gate is operated conditioned on no-decay. The absence of actual spontaneous emission events does not indicate a lack of interaction with vacuum fluctuations. Rather, one can regard the conditional no-decay as a disguised form of post-selection via measurement of an open quantum system. A natural consequence is that the wave function remains a pure state, but its dynamics is influenced by decay channels, as exhibited in Fig. \ref{fig:gate_error_scan}. Successful operations of quantum coherent control generally rule out the presence of spontaneous emission events, but the discussions so far demonstrate that the design of quantum coherent control still needs to consider decay as part of the system. 

In principle, a perfect and complete conditional no-decay is practically impossible, because it requires a perfect monitoring device that collects all the spontaneous emission events with 100\% efficiency. More profoundly, for quantum logic gates, conditioning or post-selection originates from QEC, and in particular from the error detection of QEC. Consequently, we discuss modifications to typical MCWF calculations compatible with realistic imperfect monitoring and conditioning of no-decay events with emphasis on two typical situations. The first situation is that decays are monitored continuously, but the detector is not ideal, such as a practical single-photon counting module (SPCM). For a specific quantum trajectory over a given time interval $\Delta t$, the trajectory is terminated if the calculation indicates that the detector flags a decay event. There are two types of errors: first, that an actual decay event occurs but the detector misses it with probability $\eta_0(\Delta t)$, and second, that the detector responds but no actual decay event occurs, with probability $\eta_1(\Delta t)$. Generate an independent pseudo-random number $\xi \in [0, 1]$. If a quantum jump due to a decay event occurs, then the trajectory is stopped if $\xi >\eta_0(\Delta t)$; otherwise, the calculation continues. If no quantum jump occurs, then the calculation continues only if $\xi > \eta_1(\Delta t)$. 

In the second typical situation, spontaneous emission is not monitored during the interaction, but post-selection is enforced based on measurement of the qubit states at the end. The ground level of an atom usually contains a cluster of many states, and when spontaneous emission occurs there is a significant chance that the atom decays into states other than $|0\rangle, |1\rangle$. Therefore, this situation translates to carrying out the typical MCWF procedure including all these states and post-selecting trajectories that end up in $|0\rangle, |1\rangle$.

Furthermore, appropriate utilization of SECOND can equip the MCWF approach with advanced functionalities for accurately evaluating statistics of decay events. This not only provides an insightful tool for studying waiting-time distributions \cite{RevModPhys.70.101, Brandes2008WaitingTime, PhysRevB.104.195408, PhysRevB.110.094315}, but also offers essential help in revealing the physics of quantum jumps' probabilistic properties \cite{PhysRevLett.57.1696}.

The basic mechanism can be explained by calculating the probability $p_0([0, t])$ of no decay event in $[0, t]$ in a recursive way. Given the value of $p_0([0, t])$ for a system governed by unitary evolution $H_\text{U}$ and decays $H_\text{D}$, suppose the conditional no-decay wave function $\mathbf{C}(t)$ is obtained by solving SECOND as in Eq. \eqref{eq:SECOND_general}, then the probability of receiving an initial decay event in channel $e_m$ during $[t, t+dt]$ is $p_0([0, t])\gamma_{e_m}|C_{e_m}|^2dt$. Consequently, denote $p_0([0, t])$ as a real-valued function $p_0(t)$, according to the Total Probability Theorem,
\begin{equation}
\label{eq:conditional_nodecay_prob}
p_0(t+dt) = p_0(t)\big(1 - \sum_{m=1}^{M}\gamma_{e_m}|C_{e_m}(t)|^2 dt\big),
\end{equation}
which yields $\dot{p_0}(t) = \sum_{m=1}^{M}\gamma_{e_m}|C_{e_m}(t)|^2 p_0(t)$ in differential equation form with solution of $p_0(t)=\exp(\int_0^t \sum_{m=1}^{M}\gamma_{e_m}|C_{e_m}(t')|^2 dt')$. With methods in this category, the probability of a specific distribution of decay events can then be explicitly calculated. Notably, $\int_0^t p_0(t')\gamma_{e_m}|C_{e_m}(t')|^2dt'$ yields the probability of obtaining at least one event in decay channel $m$. In other words, SECOND enables the precise tracing of decay events under various prescribed conditions via MCWF. 

The spontaneous emission does not only involve the gate but also the readout processes. The readout of qubit states occupies a vital role in quantum computing and Eq. \eqref{eq:conditional_nodecay_prob} supports the investigation of how the spontaneous emission fundamentally affects the readout fidelity. For atomic qubits, the readout process \cite{RevModPhys.82.2313, PhysRevLett.114.100503, PhysRevLett.119.180503, PhysRevLett.119.180504, PhysRevLett.121.240501, PhysRevLett.134.240802, Infleqtion2025PRXquantum} can be interpreted as only coupling $|1\rangle$ to an excited state $|e\rangle$ with ample spontaneous emission rates, such that detection of spontaneous emission indicates the qubit projected onto $|1\rangle$ and otherwise $|0\rangle$. For the dynamics of readout process, detecting $|1\rangle$ conforms to conditional no-decay evolution for some finite time until the first spontaneous emission event flares, while detecting $|0\rangle$ means the observation of strict no-decay evolution. Namely, at time $t$ such a practical instantiation will determine the population in $|0\rangle$ as $p_0(t)$. 

\begin{figure}[h]
\centering
\includegraphics[width=0.4\textwidth]{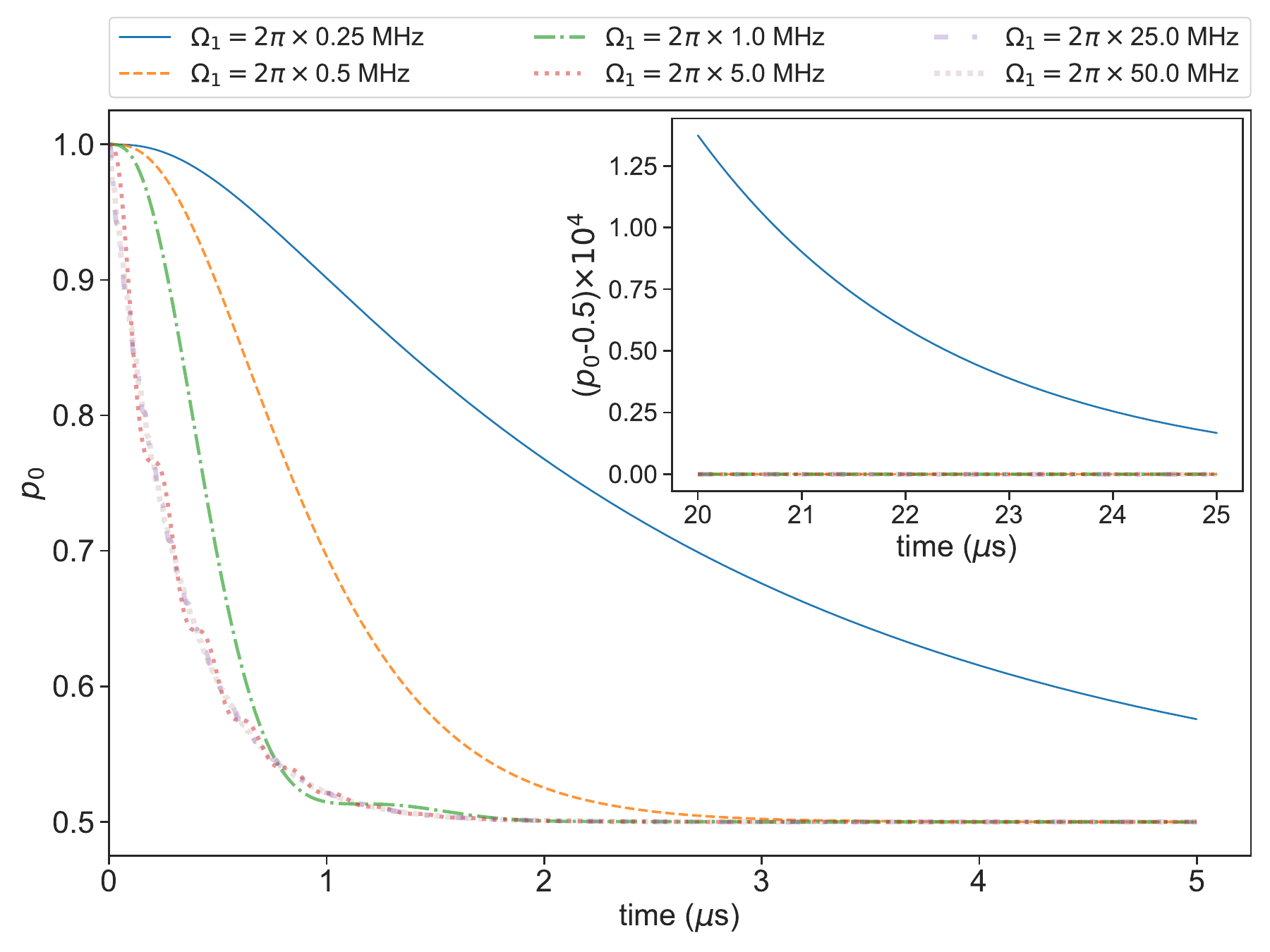}
\caption{(Color online) No-decay probability $p_0(t)$ for different Rabi frequencies in the readout process. The inset shows the situation at longer times. Here $\Delta=0, \gamma=2\pi\times 1\text{ MHz}$.}
\label{fig:p0_curves}
\end{figure}

More specifically, according to Eq. \eqref{eq:SECOND_general}, for the three-level model $|0\rangle, |1\rangle, |e\rangle$ of readout, the time evolution conditioned on no-decay corresponds to $\mathcal{H}_3/\hbar$ as in Eq. \eqref{eq:Raman_H3} but with $\Omega_0=0$. Then $p_0(t)$ can be calculated by numerically solving SECOND and integrating Eq. \eqref{eq:conditional_nodecay_prob}. For example, suppose the initial state to be read as $(|0\rangle + |1\rangle)/\sqrt{2}$ and the result is shown in Fig. \ref{fig:p0_curves}. While the spontaneous emission rate $\gamma_e$ sets the ultimate readout speed, weak readout laser necessitates extraordinarily long time cost for the qubit to fully project onto basis. This establishes a fundamental limit for the design of fast high-fidelity readout process. The Supplemental Material includes another example of no-decay probability calculation for the CZ gate.

In conclusion, we have developed a special form of wave function approach conditioned on no-decay and have applied it to analyze and refine the quantum logic gate and readout processes with respect to the influence of spontaneous emission. The prerequisite of conditional no-decay arises naturally in quantum computing as decay is generally recognized as error. The design of high-fidelity gates needs to consider these subtle effects, while population-transfer gates and purely phase gates experience distinct influences. For atomic qubits, as the fidelity of single-qubit gates and Rydberg blockade entangling gates continues to improve, the issues of spontaneous emission need to be carefully addressed both in manipulating physical qubits and implementing QEC. We anticipate that the results of this work will facilitate in-depth analysis of relevant experimental observables and will be helpful in quantum computing and quantum precision measurement.


The author gratefully acknowledges support from the National Key R\&D Program of China (Grant No. 2024YFB4504002), the Science and Technology Commission of Shanghai Municipality (Grant No. 24DP2600202), and the National Natural Science Foundation of China (Grant No. 92165107). 

\bibliographystyle{apsrev4-2}

\renewcommand{\baselinestretch}{1}
\normalsize

\bibliography{alchemy_ref}
\end{document}


\preprint{}

\title{Supplemental Material: \\ Refine coherent control of atomic qubits via wave-function approach conditioned on no-decay}

\author{Yuan Sun}
\email[email: ]{yuansun@siom.ac.cn}
\affiliation{CAS Key Laboratory of Quantum Optics and Center of Cold Atom Physics, Shanghai Institute of Optics and Fine Mechanics, Chinese Academy of Sciences, Shanghai 201800, China}
\affiliation{University of Chinese Academy of Sciences, Beijing 100049, China}

\date{\today}

\maketitle

This supplementary material contains the following contents. (\RN{1}) More details and examples about calculating the time evolution of qubits via the wave function approach conditioned on no decay. (\RN{2}) More details and examples about analyzing statistical properties associated with the time evolution of qubits via the wave function approach conditioned on no decay.

\section{Time evolution via wave function approach conditioned on no-decay}
\label{sec:SECOND_evolution}

\begin{figure}[h]
\centering
\includegraphics[width=0.6\textwidth]{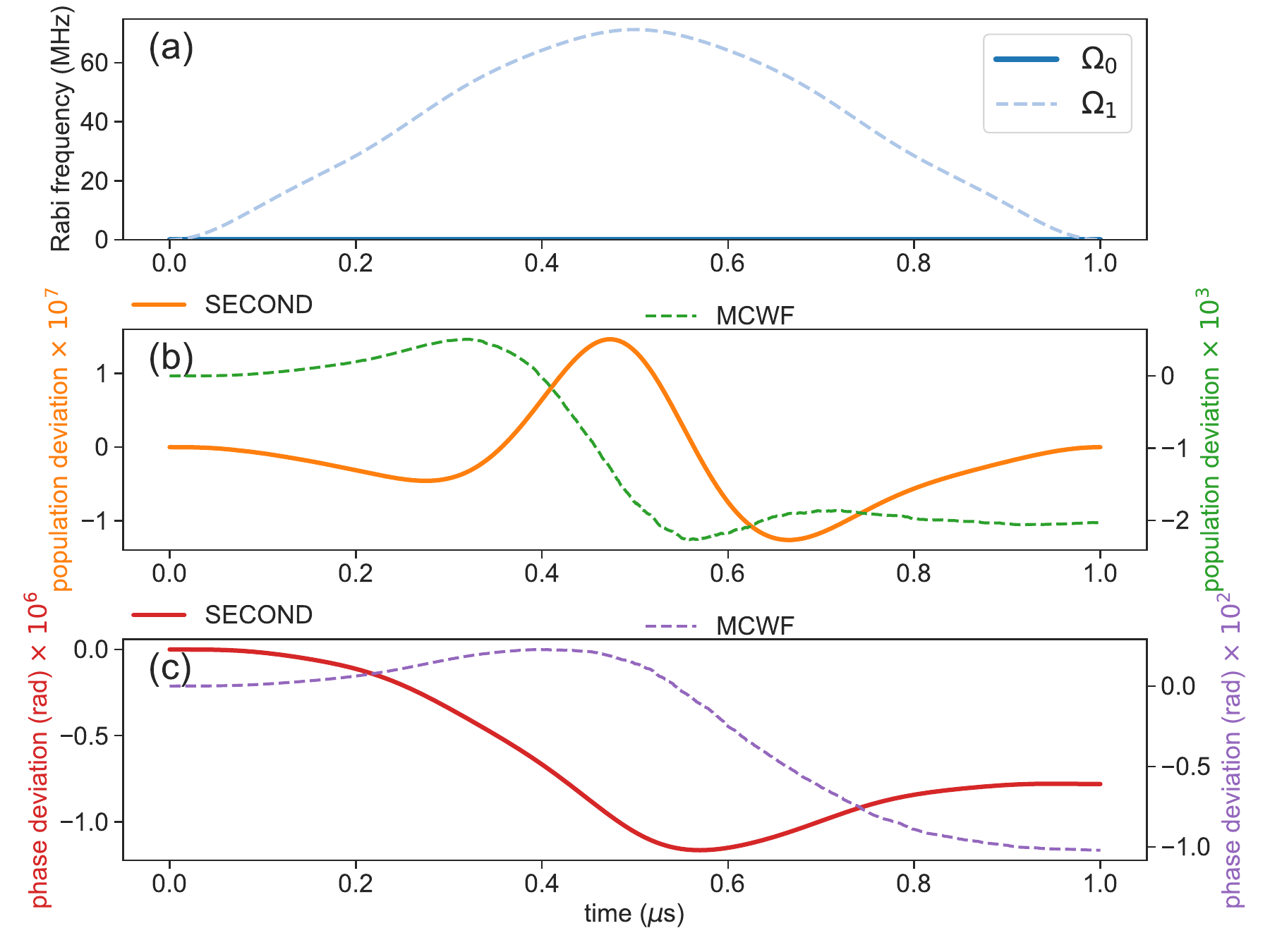}
\caption{(Color online) Comparisons between unitary time evolution ($U$), SECOND ($S$) and MCWF ($M$, averaged over 250000 trajectories). (a) Waveform of $\Omega_1(t)$. (b) Population differences $P_S - P_U$ and $P_M - P_U$ in state $|1\rangle$. (c) Phase differences $\phi_S - \phi_U$ and $\phi_M - \phi_U$ in state $|1\rangle$.}
\label{fig:phase_gate}
\end{figure}

The main text demonstrates the calculation of a single-qubit gate driven by a $\frac{\pi}{2}$-pulse in a three-level atom model with Raman transition, and it belongs to the category of population-transfer gate. The single-qubit gate also includes the category of phase gate to induce a finite phase change on the qubit wave function. Essentially, such a process and similar variations utilize the ac Stark shift to generate and accumulate appropriate phase changes. For a specific example, a $\pi$ phase shift on $|1\rangle$ will be considered here. As stated in the main text, the system consists of three atomic states $|0\rangle, |1\rangle, |e\rangle$ and description of the process conforms to Eq. (6) in the main text with $\Omega_0=0$. Fig. \ref{fig:phase_gate} presents the results of numerical calculations. As in the main text, $\gamma=2\pi\times 1\text{ MHz}$ and the Monte-Carlo wave function (MCWF) calculation assumes that $|e\rangle$ decays to $|0\rangle, |1\rangle$ with equal probabilities.

The explicit forms of the equations of motion are:

\begin{equation}
i\frac{d}{dt} \begin{bmatrix}C_0\\C_1\\C_e \end{bmatrix} = 
\begin{bmatrix} i\frac{\gamma}{2}|C_e|^2 & 0 & \frac{\Omega_0(t)}{2}\\ 
0 & \delta(t) + i\frac{\gamma}{2}|C_e|^2 & \frac{\Omega_1(t)}{2}\\ 
\frac{\Omega_0(t)}{2} & \frac{\Omega_1(t)}{2} & \Delta(t) -i\frac{\gamma}{2} +i\frac{\gamma}{2}|C_e|^2 \end{bmatrix}
\cdot\begin{bmatrix}C_0\\C_1\\C_e \end{bmatrix},
\end{equation}
where $\Omega_0=0$ for Fig. \ref{fig:phase_gate}. As analyzed in the main text, for phase gates with non-pathological smooth modulated drivings that have zero values at the end points, the difference up to the first order in unitary time evolution and SECOND calculations is zero after the interaction, as also exhibited in Fig. \ref{fig:phase_gate}.

The main text investigates the Rydberg blockade CZ gate via the synthetic modulated driving, and these discussions are immediately applicable to both the one-photon and two-photon ground-Rydberg transitions of alkali or alkali-earth atoms \cite{nphys1183, PhysRevLett.104.010503, PhysRevA.92.022336}. Meanwhile, the concepts can also be extended to other types of physical qubits, which await further study. Here the focus is on the Hamiltonian of two-qubit system mediated by the Rydberg dipole-dipole interaction for quantum computing, which has been thoroughly discussed previously \cite{RevModPhys.82.2313, PhysRevApplied.13.024059, PhysRevApplied.15.054020, PhysRevA.105.042430, OptEx480513, Yuan2024FR2}. For clarity, the details of applying Schr\"odinger equation conditioned on no-decay (SECOND) to this system will be provided here.

Regarding the unitary time evolution Hamiltonian $H_\text{U}=H_{d1}+H_{dr}$ for the Rydberg blockade CZ gate via two-photon ground-Rydberg transition $|1\rangle\leftrightarrow|e\rangle\leftrightarrow|r\rangle$ \cite{PhysRevA.105.042430, OptEx480513}, the single-body processes associated with $|01\rangle$ or $|10\rangle$ are described by the Hamiltonian $H_{d1} = H_{d10} + H_{d01}$, while the two-body process associated with $|11\rangle$ is represented by $H_{dr} = H_{d2} + H_{dF}$ including the F\"{o}rster resonance. In the rotating wave frame after the rotating wave approximation, $H_{d10}$ outlines the dynamics of initial state $|10\rangle$:
\begin{equation}
H_{d10}/\hbar= \frac{\Omega_p}{2}|10\rangle\langle e0|+\frac{\Omega^*_p}{2}|e0\rangle\langle 10| + \frac{\Omega_S}{2}|e0\rangle\langle r0|+\frac{\Omega^*_S}{2}|r0\rangle\langle e0| + \Delta|e0\rangle\langle e0| + \delta|r0\rangle\langle r0|, 
\end{equation}
while $H_{d01}$ describes the dynamics of initial state $|01\rangle$:
\begin{equation}
H_{d01}/\hbar= \frac{\Omega_p}{2}|01\rangle\langle 0e|+\frac{\Omega^*_p}{2}|0e\rangle\langle 01| + \frac{\Omega_S}{2}|0e\rangle\langle 0r|+\frac{\Omega^*_S}{2}|0r\rangle\langle 0e| + \Delta|0e\rangle\langle 0e| + \delta|0r\rangle\langle 0r|.
\end{equation}
On the other hand, for the two-body process:
\begin{eqnarray}
H_{dr} = &\frac{\Omega_p}{\sqrt{2}}|11 \rangle \langle \tilde{e}| + \frac{\Omega^*_p}{\sqrt{2}}|\tilde{e}\rangle \langle 11| + \frac{\Omega_S}{2} |\tilde{e}\rangle\langle \tilde{r}| + \frac{\Omega^*_S}{2} |\tilde{r}\rangle\langle \tilde{e}| + \frac{\Omega_p}{2} |\tilde{r}\rangle\langle \tilde{R}| + \frac{\Omega^*_p}{2} |\tilde{R}\rangle\langle \tilde{r}| + \frac{\Omega_S}{\sqrt{2}} |\tilde{R}\rangle \langle rr| +  \frac{\Omega^*_S}{\sqrt{2}} |rr\rangle\langle\tilde{R}|  \nonumber\\
& + \Delta |\tilde{e}\rangle\langle \tilde{e}| + \delta |\tilde{r}\rangle\langle \tilde{r}|+ (\Delta+\delta) |\tilde{R}\rangle\langle \tilde{R}| + 2\delta|rr\rangle\langle rr| 
+ B|rr\rangle \langle qq'| + B|qq'\rangle \langle rr| + (2\delta + \delta_q)|qq'\rangle\langle qq'|,
\end{eqnarray}
where $|\tilde{e}\rangle = (|e1\rangle+|1e\rangle)/\sqrt{2}, |\tilde{R}\rangle = (|re\rangle+|er\rangle)/\sqrt{2}$. 

Assume the spontaneous emission rate of the intermediate state $|e\rangle$ is $\gamma_e$ and the spontaneous emission rate for Rydberg states $|r\rangle, |p\rangle, |p'\rangle$ is $\gamma_r$. Then the spontaneous emission rates for $|\tilde{e}\rangle, |\tilde{R}\rangle, |rr\rangle, |pp'\rangle$ can be evaluated as $\gamma_e, \gamma_e+\gamma_r, 2\gamma_r, 2\gamma_r$, respectively. Then $\mathcal{H}_\text{D}$ and $\mathcal{H}_\text{R}$ can be readily constructed according to the procedures established in the main text.

\section{Extra details about the statistical properties associated with decay events}
\label{sec:statistical_properties}


The main text proposes the use of SECOND to accurately evaluate statistical properties of decay events under various prescribed conditions, with the qubit readout process as a typical example. Fig. 5 of main text reveals a fundamental limit about the readout time and readout laser amplitude for a high-fidelity readout process. Furthermore, by performing the same type of calculations, we evaluate what happens at relatively strong readout laser amplitudes for atomic qubits. Fig. \ref{fig:readout_saturation} summarizes the results of such a numerical simulation. It suggests that a saturation-type behavior does exist and therefore increasing the readout laser amplitude alone cannot make the readout process arbitrarily fast, as expected. Moreover, if the readout laser amplitude is too large, then the readout fidelity may suffer from a small shift.

\begin{figure}[h]
\centering
\includegraphics[width=0.6\textwidth]{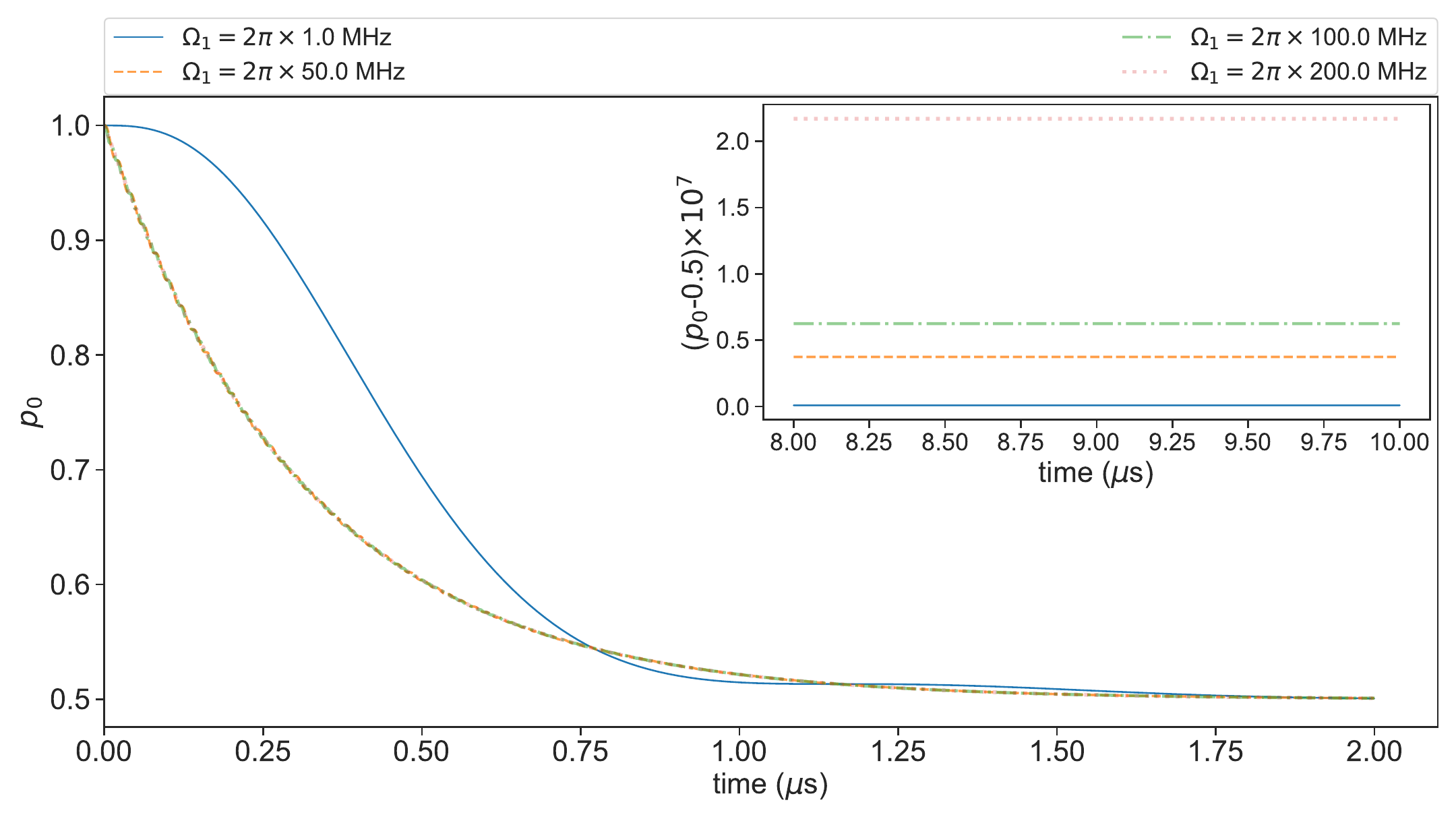}
\caption{(Color online) No-decay probability $p_0(t)$ of different $\Omega$'s for readout process as Eq. (9) in the main text, with emphasis put on the saturation behavior at large values of $\Omega$. The inset shows situation at longer times. Here $\Delta=0, \gamma=2\pi\times 1\text{ MHz}$. The initial state to be read is $(|0\rangle + |1\rangle)/\sqrt{2}$.}
\label{fig:readout_saturation}
\end{figure}


In previous experimental studies of Rydberg blockade CZ gates, the two-photon ground-Rydberg transition generally satisfies the prerequisite of adiabatic elimination. In brief, under such a hypothesis the one-photon detuning $\Delta$ is much larger than $\Omega_p, \Omega_S$, and then neglecting the spontaneous emissions of intermediate level becomes appropriate. Typically, the choice of intermediate state $|e\rangle$ meets the requirement of linewidth $\Gamma_e\lesssim 1$ MHz for $\sim 100$ ns gate time, if the goal is to control gate errors on the order of $1 \times 10^{-4}$. Now with the help of advanced functionalities for accurately evaluating statistics of decay events enabled by SECOND, it is possible to numerically calculate the precise probability of encountering spontaneous emission during the course of gate. More specifically, Fig. \ref{fig:readout_saturation} displays an example of calculating the no-decay probability $p_0(t) = p_0([0, t])$ of a typical Rydberg blockade CZ gate with the same gate parameters as Fig. 3 in the main text. In the case of $\gamma_r=0$ the spontaneous emission of the intermediate state $|e\rangle$ causes the decay with $\gamma_e = 2\pi\times 5 \text{ MHz}$.

\begin{figure}[h]
\centering
\includegraphics[width=0.6\textwidth]{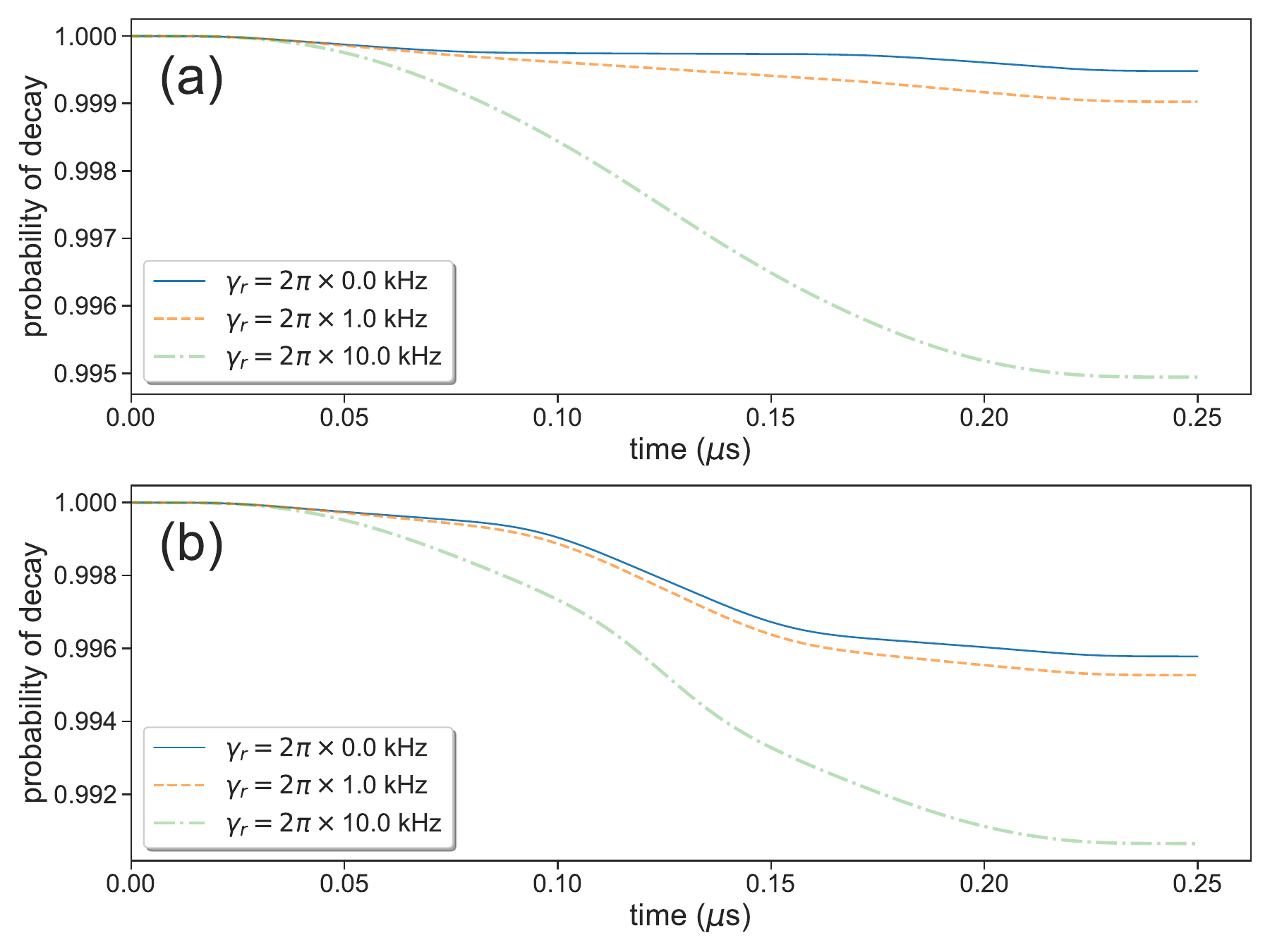}
\caption{(Color online) No-decay probability $p_0(t)$ for the Rydberg blockade CZ gate. (a) Single-body process with initial state $|01\rangle$ or $|10\rangle$. (b) Two-body process with initial state $|11\rangle$.}
\label{fig:readout_saturation}
\end{figure}


The main text outlines the modifications to typical MCWF calculations compatible with realistic imperfect monitoring and conditioning of no-decay events. Here, we provide more details on this using the two-level atom model. As mentioned in Eq. (4) of the main text, when the monitoring of spontaneous emission is ideal, then the wave function $(X, Y)$ obeys SECOND in the following form:
\begin{equation}
\label{eq:SECOND_2level}
i\frac{d}{dt} 
\begin{bmatrix} X \\ Y \end{bmatrix}
= 
\begin{bmatrix} \frac{i}{2}\gamma|Y|^2 & \frac{\Omega(t)}{2}\\ 
\frac{\Omega(t)}{2} & \Delta(t)  - \frac{i}{2}\gamma + \frac{i}{2}\gamma|Y|^2 \end{bmatrix}
\cdot \begin{bmatrix} X \\ Y \end{bmatrix},
\end{equation}
where $\gamma$ is the decay rate of excited state and $X, Y$ represent the wave functions of ground state $|g\rangle$ and excited state $|e\rangle$ respectively.

When the decays are monitored all the time but the detection of decay events is not perfect, the time evolution conditioned on no-decay according to such detections no longer strictly conforms to Eq. \eqref{eq:SECOND_2level}. The imposed post-selection or conditioning sets the requirement that if the calculation finds a decay event in the detection result, then such a quantum trajectory will be discarded. The MCWF approach can calculate the quantum trajectory under this situation as well. More specifically, the wave function $(X_1, Y_1)$ at the next time instant $t + \Delta t$ for small time interval $\Delta t$ can be numerically deduced according to the known wave function $(X_0, Y_0)$ at time $t$. Suppose that an actual decay event occurs but the detector misses it with probability $\eta_0(\Delta t)$ and that the detector reports a decay event falsely with probability $\eta_1(\Delta t)$. Then the evaluation procedure contains the following steps. (A) The chance of having a spontaneous emission from $|e\rangle$ over $[t, t+\Delta t]$ can be approximated by $\gamma |Y_0|^2\Delta t$. Get the values of two independent pseudo random numbers $\xi, \eta \in [0, 1]$, if $\xi > \gamma |Y_0|^2\Delta t$ then perform step (B), otherwise perform step (C). (B) This emulates that a decay event actually occurs. If $\eta > \eta_0(\Delta t)$, it indicates the detection catches the decay event correctly and then the calculation stops. Otherwise, the calculation continues with $(X_1, Y_1)=(1, 0)$ induced by the decay. (C) This emulates that a decay event does not occur. If $\eta < \eta_1(\Delta t)$, it indicates the detection falsely reports a decay and then the calculation stops. Otherwise, $(X_1, Y_1)$ can be obtained by solving Eq. \eqref{eq:SECOND_2level} over $[t, t+\Delta t]$ with the initial condition $(X_0, Y_0)$.


\bibliography{alchemy_ref}